\documentclass[aps,prb,
showpacs]{revtex4}

\usepackage{amsmath}
\usepackage{color}
\usepackage{amssymb}

\usepackage{graphicx}

\begin{document}

\title{Antiferromagnetic order in CeCoIn$_5$ oriented by spin-orbital coupling}

\author{V.P.Mineev$^{1,2}$}
\affiliation{$^1$
 Commissariat a l'Energie Atomique, UGA, INAC-FELIQS, 38000 Grenoble, France;\\
$^2~$Landau Institute for Theoretical Physics, 119334, Moscow, Russia}

\begin{abstract}
An incommensurate spin density wave ($Q$ phase) confined inside the superconducting state 
at high basal plane magnetic field is an unique  property of the heavy fermion metal CeCoIn$_5$. The neutron scattering
experiments and the theoretical studies point out that this state come out from the soft mode  condensation
of magnetic resonance excitations.
We show that the fixation of direction of antiferromagnetic modulations by a magnetic field 
reported by Gerber et al., Nat. Phys. {\bf 10}, 126 (2014) is explained by spin-orbit coupling.
This result, obtained on the basis of quite general phenomenological arguments, is supported 
by the microscopic  derivation of  the  $\chi_{zz}$ susceptibility dependence on the 
mutual orientation of the basal plane magnetic field and the direction of modulation of spin   polarization
 in  a multi-band metal.
\end{abstract}

\date{\today}
\maketitle

\section{Introduction}

CeCoIn$_5$ is a tetragonal, d-wave-pairing superconductor with the highest critical temperature $T_c=2.3 K$
among all the heavy fermion compounds \cite{Petrovic, Izawa,Vorontsov}.
The superconducting state of CeCoIn$_5$ at a magnetic field above 9.8 T applied in the basal plane (Fig.1) of its tetragonal crystal structure co-exists with incommensurate antiferromagnetic (AF) ordering  or spin density wave (SDW)  \cite{Kenzel} 
with ${\bf Q}_{IC}=(0.45,\pm 0.45, 0.5)$ independent of the field magnitude.
Its 2D incommensurate part ${\bf q}_{IC}=(0.45,\pm 0.45, 0)$ is parallel to the nodal directions of the $d$-wave order parameter 
$$
\Delta({\bf k})=\Delta(\cos k_x^2-\cos k_y^2).
$$ 
Here we use reciprocal lattice units.  The existence of the magnetic order was first detected by the technique of NMR \cite{Young} and its precise field-dependence later determined \cite{Koutroulakis}.
The antiferromagnetic modulation is concentrated 
 on the  Cerium sites  with amplitude $m=0.15\mu_B$ ($\mu_B$ is the Bohr magneton) and polarized along the tetragonal axis. The incommensurate SDW is confined inside the superconducting phase, meaning that here superconductivity is an essential ingredient for SDW to develop.
Different  theoretical models have been proposed to explain why the SDW order occurs only in the high field-low temperature region of the $d$-wave superconducting state
\cite{Miyake,Yanase, Agterberg,Aperis,Ikeda2010,Ikeda2011,Machida,Kato,Michal}. The  choice  between these  models can be made with the help
of results of neutron scattering. 

In the zero-field  superconducting state,
a spin resonance   was found at  a frequency $\omega$=0.6 mev $ \approx$ 7K,  which  corresponds approximately to $3T_c$. Initially, the resonance  was observed \cite{Stock} at a wave vector ${\bf Q}=(0.5,0.5, 0.5)$  associated with the nested parts of the Fermi surface corresponding to antiferromagnetic correlations. Theoretically Eremin and collaborators\cite{Eremin} have attributed the resonance to the proximity to the threshold of the particle-hole excitations continuum which is at energy $\omega_c=\min(|\Delta_{\bf k}|+|\Delta_{{\bf k}+{\bf Q}}|)$. Another scenario related to a magnon excitation was proposed by Chubukov and Gor'kov\cite{Chubukov}. 

Recent, more precise inelastic neutron scattering measurements \cite{Raymond2015} have demonstrated that the spin resonance is peaked  at the same wave vector as  the incommensurate static AF  modulation in a high fields. Moreover, the fluctuations associated with the resonance are polarized along the tetragonal c-axis that corresponds to the direction of the ordered magnetic moments in the $ Q$ phase. Thus, the dynamical mode at zero field and the field induced static order share the same properties as  if the resonance is a dynamical precursor of the $Q$ phase. Also there were observed  the Zeeman splitting of the resonance under magnetic field 
and the 
 softening of the lowest energy mode of the Zeeman-split resonance \cite{Panarin, Broholm,Raymond2012}.
All these observations 
 point on the soft mode condensation  caused by magnetic field as the mechanism for AF ordering formation. Such type 
theoretical model
 was  put forward in Ref.16.
 
  In a Random Phase Approximation the spin susceptibility is
  $$
  \chi(\mathbf{q},\omega)=\frac{\chi^{(0)}(\mathbf{q},\omega)}{1-U_q\chi^{(0)}(\mathbf{q},\omega)},
  $$ 
  where $\chi^{(0)}(\mathbf{q},\omega))$ is the electron gas  susceptibility in superconducting  state and $U_q$ is a momentum-dependent Hubbard-Coulomb repulsion potential. In this model the conditions for a collective excitation (called spin exciton) to occur are $U_q\Re\mathfrak{e}\chi^{(0)}(\mathbf{q},\omega)=1$, and $U_q\Im\mathfrak{m}\chi^{(0)}(\mathbf{q},\omega)\ll1$.
The static antiferromagnetic  state  at some ${\bf q}$ is realized 
when the real part of  the spin susceptibility along the tetragonal $c$-axis in the presence of a finite basal plane magnetic field ${\bf H}_\perp$ exceeds  the inverse  constant of the AF interaction
\begin{equation}
\Re e\chi^{(0)}_{zz}(\omega=0,{\bf q}, H_{\perp})>U^{-1}_{{\bf q}}.
\end{equation}
In a two-dimensional model (Ref.16) it was found that  $\chi^{(0)}_{zz}(\omega=0,{\bf q}_{IC}, H_{\perp})$ in $d$-wave superconducting state  increases with the field and exceeds the corresponding normal state susceptibility at  fields essentially smaller than the paramagnetic limiting field \cite{footnote}.  The physical reason for this behavior is that the  incommensurate wave vector connects the points on the Fermi surface where $\Delta({\bf k})=-\Delta({\bf k}+{\bf q}_{IC})$. The same is true in real 3D case with  modulation along ${\bf Q}_{IC}$.
As a result, the tendency for AF instability  in CeCoIn$_5$ is much more effective  in the superconducting $d$-wave state.

An important observation made 
 recently \cite{Gerber} is that  the degeneracy  between the two possible directions 
of antiferromagnetic modulation ${\bf Q}_{IC}=(0.45,\pm 0.45, 0.5)$ is lifted by the magnetic field orientation. 
Namely, for field parallel to $[1\bar 1 0]$, a Bragg peak  was with ${\bf Q}_h=(0.45,0.45, 0.5)$ but no peaks corresponding to 
${\bf Q}_v=(0.45,-0.45, 0.5)$ were detected.
For field precisely parallel to [100], the direction of  the incommensurate part of AF modulation can take either of the two directions parallel to the gap nodes of $d_{x^2-y^2}$ pairing. But a tiny deviation of  the field orientation from [100] toward [110]  lifts this degeneracy and fixes the AF modulation along [the 1${\bar 1}$0] direction.
Correspondingly, a deviation of the field  from [100] toward  [1${\bar 1}$0] fixes the AF modulation along [110] direction.
In other words, the incommensurate AF modulation chooses that orientation  which is the "most perpendicular" to the field direction (see Fig.2).

The authors of Ref.25 have attributed this phenomenon to the presence in
the Q phase of an additional modulated on atomic scale superconducting component 
 with triplet $p$-pairing  with order parameter $\hat\Delta_Q=i({\bf d}_Q\mbox{\boldmath$\sigma$})\sigma_y$ called pair density wave (PDW) interacting with $d$-wave order $\Delta$ and the AF order $M^z_Q$
 $$
V\propto iM^z_Q\left (\Delta^\dagger d^z_{-Q}-\Delta(d^z_Q)^\dagger\right)+c.c.
$$
For the ${\bf H} \parallel [1\bar 10]$  the$p$-wave state with maximal spin-susceptibility along field direction and the zeros of the order parameter along perpendicular to the field  direction ($d_Q^z\propto(k_x-k_y))$ does not disturb the d-wave superconducting state and can stimulate the  emergence of $Q$-state with ${\bf Q}=(0.45,0.45,0.5)$.
While possible phenomenologically this idea was not supported by any argument in favor of a specific mechanism for space modulated triplet pairing in this material.

 Another interpretation of the same phenomenon,   developed quite recently \cite{Ikeda2015}, is based on the AF modulation interaction with the Fulde-Ferrel-Larkin-Ovchinnikov  (FFLO) modulation  parallel to the magnetic field direction. One can remark, however, that the phase diagram for the coexistence of  the superconducting $Q$-state and the  FFLO state found theoretically in the previous paper of the same authors \cite{Ikeda2011} does not resemble on the phase diagram of superconducting $Q$-state shown on Fig.1. Moreover, the isothermal measurements 
 \cite{Tokiwa2012}
 at $H\parallel [100]$  did not reveal an entropy increase at phase transition from the superconducting state mixed state to the  superconducting $Q$-state which would indicate nodal quasiparticles in FFLO SC state. By
contrast, 
 a clear
reduction of the entropy
 is found at
a second-order  transition at 10.4 T. This transition
coincides with the  incommensurate
AF order \cite{Young,Kenzel}. The observed reduction of DOS is in
perfect agreement with the expectation for a SDW formation
without the additional FFLO state.

 Here we propose an explanation of the $Q$-phase anisotropy governed by magnetic field  based not on an imaginary additional ordering, but arising from  the ordinary spin-orbit coupling.
In the next Section,
on the basis of quite general phenomenological arguments, we demonstrate, that 
in a tetragonal crystal
under the basal plane magnetic field,  the static magnetic susceptibility along the tetragonal axis $\chi^{(0)}_{zz}({\bf q}_\perp,{\bf H}_{\perp})$ 
at finite ${\bf q}_\perp=(q_x,q_y)$
is largest either at  
${\bf q}_\perp\parallel {\bf H}_{\perp}$ or at ${\bf q}_\perp\perp {\bf H}_{\perp}$.  As a result,  if  the maximum of the susceptibility occurs for a   space modulation perpendicular to the field, then  the inequality given by Eq.(1)  is  realized first on the ${\bf Q}_{IC}$ direction, which is closer to being perpendicular to the magnetic-field direction. Thus,
the degeneracy of directions of antiferromagnetic instability  is lifted. 

The $\chi^{(0)}_{zz}(\omega=0,{\bf q}, H_{\perp})$ calculated in a single band model \cite{Michal} is completely independent of  the mutual orientation of ${\bf q}$ and the basal plane magnetic field ${\bf H}_{\perp}$. One can show that this orientational independence persists  in a single band metal even  in the Abrikosov mixed state characterized by  inhomogeneous superfluid velocity and field distributions. 
On the other hand, it is known that the spin-orbit interaction  in a non-centrosymmetric tetragonal metal causes the magnetic susceptibility 
 orthorhombic anisotropy \cite{Takimoto}  $\chi_{xx}-\chi_{yy}\sim q_x^2-q_y^2$. Such type anomalous susceptibility anisotropy in noncenrosymmetric CePt$_3$Si  has been mesured recently
by polarized neutron scattering  reported in  Ref.27. 
A similar phenomenon can be expected in a multi-band  centrosymmetric material. 
CeCoIn$_5$ is the multi-band metal that has been  established by  the de Haas-van Alphen measurements \cite{Settai},  by the  band structure calculations  and  by the  band spectroscopy studies \cite{Tanaka, Yazdani2012, Davis,Yazdani2013}.
 In the third section  to illustrate the phenomenological conclusions we show 
that, owing  to  the  interband spin-orbit interaction,
the  static spin susceptibility $\chi^{(0)}_{zz}({\bf q}_\perp,{\bf H}_{\perp})$ in a tetragonal multi-band metal  depends on the  mutual orientation
of the wave vector ${\bf q}_\perp$ and the magnetic field ${\bf H}_{\perp}$.
The spin-orbit coupling originating from the interaction of conducting electrons with the ionic  crystal field  can in principle be another source  of violation of tetragonal symmetry by the basal plane magnetic field.

The presented microscopic calculations of susceptibility are performed  for a two-band tetragonal metal in the normal state.
However, it should be stressed that 
antiferromagnetism  and antiferromagnetic domain switching are phenomena originating from  different 
mechanisms.  As it was demonstrated in Ref.16, the antiferromagnetism arises from an  anomalous enhancement of  the $\chi^{(0)}_{zz}({\bf q}_{IC})$  susceptibility in the $d$-wave superconducting state under the basal plane magnetic field.  The domain switching is caused by spin-orbit coupling violating the crystal  tetragonal symmetry at finite basal plane magnetic field and space modulation. 
The susceptibility $\chi^{(0)}_{zz}({\bf q}_\perp,{\bf H}_{\perp})$ is proved to be dependent from the mutual orientation of 
the magnetic field and the direction of the  space modulated spin polarization.  This dependence takes place already in
the CeCoIn$_5$ normal state,  but reveals itself  in the $d$-wave superconducting state where the antiferromagnetic modulation developes. To demonstrate the violation of tetragonal symmetry by the basal plane magnetic field, it is sufficient  to perform a microscopic  derivation of the susceptibility in the normal state. 
  The corresponding calculation in the superconducting state is much more cumbersome, but does not add any qualitative changes to the conclusions based on
   the  normal state calculations.

\section{ Phenomenological approach}

We consider a tetragonal paramagnet in a magnetic field  having a constant basal plane  $(x,y)$ part and coordinate-dependent small additions
\begin{equation}
{\bf H}({\bf r})=[H_x+\delta H_x({\bf r})]\hat x+[H_y+\delta H_y({\bf r})]\hat y+\delta H_z({\bf r})\hat z.
\end{equation}
Its  free energy in the quadratic approximation has the following form
\begin{widetext}
\begin{eqnarray}
{\cal F}=\int dV\left \{\alpha(M_x^2+M_y^2)+\alpha_zM_z^2 +\gamma\left [\left (\frac{\partial{\bf M}}{\partial x } \right )^2+\left (\frac{\partial{\bf M}}{\partial y } \right )^2\right ]+\gamma_z\left (\frac{\partial{\bf M}}{\partial z } \right )^2+\gamma_{xy}\frac{\partial M_x}{\partial x }\frac{\partial M_y}{\partial y }\right.~~~~~~~~~~~~~~~~~~~~\nonumber\\
+\gamma_{\perp z}\left (\frac{\partial M_x}{\partial x }+\frac{\partial M_y}{\partial y } \right )\frac{\partial M_z}{\partial z }\left.+\delta_1\left ( H_x\frac{\partial M_z}{\partial y}-H_y\frac{\partial M_z}{\partial x} \right )^2+    \delta_2\left (H_x\frac{\partial M_z}{\partial x}+H_y\frac{\partial M_z}{\partial y}   \right )^2-{\bf H}{\bf M} \right \}.~~~~~~~~~~~~~~~~~~~~
\label{fe}
\end{eqnarray}
\end{widetext}
Here, all terms besides the terms proportional to $\delta_1$ and $\delta_2$ have  the tetragonal symmetry, whereas these two terms 
depend on the   mutual orientation of the basal plane field and the direction of the space modulation of the $M_z$ component of magnetization. This dependence originates from the spin-orbit interaction. Disregarding the   spin-orbit coupling corresponds to the equality $\delta_1=\delta_2$,  recreating the functional tetragonal symmetry.  For
$\delta_1>\delta_2$, the direction of the $M_z$ basal-plane modulation is preferentially   parallel to $$ {\bf H}_\perp=H_x\hat x+H_y\hat y.$$ On the other hand, for $\delta_1<\delta_2$, the direction of the $M_z$ basal-plane modulation tends to be  perpendicular  to $ {\bf H}_\perp$.  

One can formulate the same conclusion in terms of the susceptibility. Namely, by making the variation of Eq.(\ref{fe}) with respect to the
magnetization components,  taking into account the expressions for the equilibrium parts of the magnetization 
$$M_x=H_x/2\alpha,~~~~~~M_y=H_y/2\alpha,$$ 
and performing a Fourier transform, one arrives that  the following equations for the magnetization response to the  coordinate-dependent part of the magnetic field
$\delta{\bf H}({\bf r})$:
\begin{widetext}
\begin{eqnarray}
\label{M_x}
2\left [\alpha+\gamma(q_x^2+q_y^2)+\gamma_zq_z^2\right]\delta M_x+\gamma_{xy}k_xk_y\delta M_y+\gamma_{\perp z}q_xq_z\delta M_z=\delta H_x,\\
\label{M}
\gamma_{xy}k_xk_y\delta M_x+2\left [\alpha+\gamma(q_x^2+q_y^2)+\gamma_zq_z^2\right]\delta M_y++\gamma_{\perp z}q_yq_z\delta M_z=\delta H_y,\\
\label{M_z}
\gamma_{\perp z}(q_xq_z\delta M_x+q_yq_z\delta M_y)+2\left [\alpha_z+\gamma(q_x^2+q_y^2)+\gamma_zq_z^2
+\delta_1(H_xq_y-H_yq_x)^2+\delta_2(H_xq_x+H_yq_y)^2\right ]\delta M_z=\delta H_z.
\end{eqnarray}
\end{widetext}
The solution of these equations  yields the Fourier components of magnetization,
\begin{eqnarray}
\delta M_x({\bf q})=\chi_{xx}\delta H_x({\bf q})+\chi_{xy}\delta H_y({\bf q})+\chi_{xz}\delta H_z({\bf q}),\\
\delta M_y({\bf q})=\chi_{yx}\delta H_x({\bf q})+\chi_{yy}\delta H_y({\bf q})+\chi_{yz}\delta H_z({\bf q}),\\
\delta M_z({\bf q})=\chi_{zx}\delta H_x({\bf q})+\chi_{zy}\delta H_y({\bf q})+\chi_{zz}\delta H_z({\bf q}).
\end{eqnarray}
The coefficients $\gamma_{xy}$, and $\gamma_{\perp z}$ have relativistic smallness relative to the exchange-determined coefficients $\alpha$ and  $\gamma$. Neglecting the entanglement between the components of magnetization in Eqs.(\ref{M_x})-(\ref{M_z}), which gives only 
small corrections of order 
$O(\gamma_{xy}^2)$ and $O(\gamma_{xy}\gamma_{\perp z})$ to the susceptibilities, we obtain
\begin{widetext}
\begin{equation}
\chi_{zz}({\bf q})\cong\frac{1}{2\left [\alpha_z+\gamma(q_x^2+q_y^2)+\gamma_zq_z^2
+\delta_1(H_xq_y-H_yq_x)^2+\delta_2(H_xq_x+H_yq_y)^2\right ] }.
\end{equation}
\end{widetext}
Thus, for $\delta_1>\delta_2$ the magnetic susceptibility along the tetragonal axis 
 is  largest for ${\bf H}_\perp||{\bf q}_\perp$, where 
 $$
 {\bf q}_\perp=(q_x,q_y).
 $$ 
 On the other hand, for  $\delta_1<\delta_2$, the perpendicular mutual orientation of ${\bf H}_\perp$ and ${\bf q}_\perp$ corresponds to a maximum in the  $z$ component of susceptibility. This conclusion becomes  evident if we rewrite the susceptibility as
 \begin{widetext}
\begin{equation}
\chi_{zz}({\bf q})\cong\frac{1}{2\left [\alpha_z+\gamma {\bf q}_\perp^2+\gamma_zq_z^2
+(\delta_1-\delta_2)[{\bf H}_\perp\times{\bf q}_\perp]^2+\delta_2{\bf H}_\perp^2{\bf q}_\perp^2\right ] }.
\end{equation}
\end{widetext}

 The derivation presented here 
 demonstrates the dependence of the susceptibility on the 
mutual orientation of the field and the direction of modulation in the long wave limit. The effect, however, can be strong enough in the case of atomic scale antiferromagnetic orderings.
 
Let us now show that the established $\chi_{zz}$
susceptibility dependence from $[{\bf H}_\perp\times{\bf q}_\perp]^2$ really takes place in a two band tetragonal metal.

\section{Microscopic derivation}

The Green function of a tetragonal two-band metal in an external magnetic field ${\bf H}_\perp$ satisfies the equation
\begin{equation}
\hat H\hat G=\hat 1
\end{equation}
with  the $4\times4$ matrix Hamiltonian
\begin{equation}
\hat H=\left( \begin{array}{cc}(i\omega-\xi_1)\sigma_0+{\bf h}_\perp\mbox{\boldmath$\sigma$}&i{\bf l}\mbox{\boldmath$\sigma$}\\
-i{\bf l}\mbox{\boldmath$\sigma$}& (i\omega-\xi_2)\sigma_0+{\bf h}_\perp\mbox{\boldmath$\sigma$}
\label{matrix}
\end{array}\right ),
\end{equation}
where $$\xi_i({\bf k})=\varepsilon_i({\bf k})-\mu, ~~~i=1,2$$ are the band energies counted from the chemical potential,  ${\bf h}_\perp=\mu_B{\bf H}_\perp$, $\mu_B$ is the Bohr magneton, and $\mbox{\boldmath$\sigma$}=(\sigma_x,\sigma_y,\sigma_z)$ are the Pauli matrices. The interband spin-orbit coupling \cite{Samokhin}  is given by the vector ${\bf l}({\bf k})$ which  is an even function ${\bf l}(-{\bf k})={\bf l}({\bf k})$
subordinating all the symmetry operations $g$ of the tetragonal point group $g{\bf l}(g^{-1}{\bf k})={\bf l}({\bf k})$. For the sake of concreteness, we can choose it to have the following form
\begin{equation}
{\bf l}({\bf k})=\gamma_\perp k_z( k_y\hat x- k_x\hat y)+\gamma_\parallel  \hat k_x \hat k_y(\hat k_x^2-\hat k_y^2)\hat z,
\end{equation}
where $\hat k_x=k_x/k_F,~\hat k_y=k_y/k_F$ such that $\gamma_\perp$ and $\gamma_\parallel$ have common dimensionality  of inverse mass $[1/m]$.
We are seeking a ${\bf q}_\perp$ dependent $z$ component of the spin susceptibility given by the equation
\begin{widetext}
\begin{equation}
\chi^{(0)}_{zz}({\bf q}_\perp)=-\mu_B^2T\sum_{{\bf k},\omega_n}Tr\left( \begin{array}{cc}\sigma_z& 0\\ 
0 & \sigma_z\end{array}\right )\hat G({\bf k}+{\bf q}_\perp/2,\omega_n)\left( \begin{array}{cc}\sigma_z& 0\\ 
0 & \sigma_z\end{array}\right )\hat G({\bf k}-{\bf q}_\perp/2,\omega_n).
\end{equation}
\end{widetext}
 Calculating the trace, we rewrite  it in terms of  the products of $\hat G$ matrix elements
\begin{widetext}
\begin{equation}
\chi^{(0)}_{zz}({\bf q})=-\mu_B^2T\sum_{{\bf k},\omega_n}\sum_{i=1,j=1}^4(-1)^{i+j}
G_{ij}({\bf k}+{\bf q}_\perp/2,\omega_n)
 G_{ji}({\bf k}-{\bf q}_\perp/2,\omega_n).
 \label{susc}
\end{equation}
\end{widetext}
 In this  complete expression, we seek the terms proportional to the combination $[{\bf H}_\perp\times{\bf q}_\perp]^2$.
Let us take, for instance,   the sum of the product of  matrix elements 
\begin{widetext}   
\begin {equation}
\mu_B^2T\sum_{{\bf k},\omega_n}\left [
G_{12}({\bf k}+{\bf q}_\perp/2,\omega_n)
 G_{21}({\bf k}-{\bf q}_\perp/2,\omega_n)+G_{21}({\bf k}+{\bf q}_\perp/2,\omega_n)
 G_{12}({\bf k}-{\bf q}_\perp/2,\omega_n)\right ].
 \label{label}
\end{equation}
\end{widetext}
Among the many terms in this sum,  we will keep only the ${\bf H}_\perp$ and ${\bf q}_\perp$ mutual orientation dependent terms. They are
\begin{widetext}   
\begin {eqnarray}
-2\mu_B^2T\sum_{{\bf k},\omega_n}
\frac{[(i\omega_n-\xi_2)^2-h_\perp^2-l_z^2]_{{\bf k}+{\bf q}_\perp/2}\Re e[(h_x+ih_y)^2(il_x+l_y)^2_{{\bf k}-{\bf q}_\perp/2}]}{D({\bf k}+{\bf q}_\perp/2,\omega_n)D({\bf k}-{\bf q}_\perp/2,\omega_n)}\nonumber \\
+\frac{[(i\omega_n-\xi_2)^2-h_\perp^2-l_z^2]_{{\bf k}-{\bf q}_\perp/2}\Re e[(h_x+ih_y)^2(il_x+l_y)^2_{{\bf k}+{\bf q}_\perp/2}]}{D({\bf k}+{\bf q}_\perp/2,\omega_n)D({\bf k}-{\bf q}_\perp/2,\omega_n)}
,
\label{product}
\end{eqnarray}
\end{widetext}
where
\begin{equation}
D({\bf k},\omega_n)=[(i\omega_n-\xi_+)^2-\xi_-^2-{\bf h}_\perp^2-{\bf l}^2]^2-4[\xi_-^2{\bf h}_\perp^2+({\bf h}_\perp{\bf l})^2],
\end{equation}
and
\begin{equation}
\xi_\pm({\bf k})=\frac{\xi_1({\bf k})\pm\xi_2({\bf k})}{2}.
\end{equation}
Keeping  the ${\bf q}_\perp$ dependence only in 
$$
\Re e[(h_x+ih_y)^2(il_x+l_y)^2_{{\bf k}\pm{\bf q}_\perp/2}],
$$ 
we obtain 
\begin{widetext} 
\begin {equation}
\mu_B^2\gamma_\perp^2
T\sum_{{\bf k},\omega_n}\frac
{ k_z^2[(i\omega_n-\xi_2({\bf k}))^2-h_\perp^2-l_z^2({\bf k})]} 
{D^2({\bf k},\omega_n)}
\left \{2[{\bf h}_\perp\times{\bf q}_\perp]^2-{\bf h}_\perp^2{\bf q}_\perp^2\right \}.
\label{product1}
\end{equation}
\end{widetext}
The sum over the Matsubara frequencies is easily calculated. However, the further integration over the Brillouin zone can only be performed  numerically, taking into account the actual intraband quasiparticle spectra and the interband spin-orbital momentum dependencies.  The knowledge of the quasiparticle density of states energy dependence near the Fermi surface is also crucially important.  Analytically,  we can write a rough 
estimation for
orientation-dependent part of the susceptibility as
\begin{equation}
\chi_{zz}^{anis1}\propto\mu_B^2N_0\left (\frac{\mu_B\gamma_\perp  k_F}{\varepsilon_F^2}  \right )^2[{\bf H}_\perp\times{\bf q}_\perp]^2.
\label{1}
\end{equation}
Taking in  mind that  the modulus of the wave vector of antiferromagnetic modulation in CeCoIn$_5$  is  $|{\bf q}_\perp|\approx k_F$
one can write the estimation of the absolute value of the susceptibility anisotropy as
\begin{equation}
\chi_{zz}^{anis1}\propto\mu_B^2N_0\left (\frac{\mu_B H}{\varepsilon_F}\right )^2\left (\frac{\gamma_\perp k_F^2}{\varepsilon_F}\right )^2.
\end{equation}
CeCoIn$_5$ is the heavy fermion compound with the electron effective mass about hundred times larger than the bare electron mass \cite{Settai} $m^*\approx 100m$. This means that  the Fermi energy  is quite small and can be of the order of spin-orbit interaction $\varepsilon_F\sim\gamma_\perp k_F^2$. At the same time, the magnetic energy $\mu_B H$ in fields about 10 Tesla is about 10 Kelvin. Hence the anisotropy of susceptibility can have noticeable magnitude in comparison with the Pauli susceptibility $\chi_P=2N_0\mu_B^2$.

Parametrically similar  contributions to the orientation dependent part of the susceptibility originate from all the Green-function products in Eq. (\ref{susc}) with indices $i\ne j$. 
On the other hand  the products with $i= j$ give rise to   contributions such as
\begin{equation}
\chi_{zz}^{anis2}\propto\mu_B^2N_0\left (\frac{\mu_B\gamma_\parallel\gamma_\perp  k_F^3}{\varepsilon_F^3}  \right )^2[{\bf H}_\perp\times{\bf q}_\perp]^2\approx
\mu_B^2N_0\left (\frac{\mu_B H}{\varepsilon_F}\right )^2\left (\frac{\gamma_\parallel \gamma_\perp k_F^4}{\varepsilon_F^2}\right )^2.
\label{2}
\end{equation}

Thus, we have demonstrated by direct microscopic calculation that the violation of tetragonal symmetry by a basal plane magnetic field discussed  in previous section on a  purely phenomenological basis,  really takes place in multiband metals.

Depending on the relative value and signs of expressions (\ref{1}) and (\ref{2}),  the mutual orientations of ${\bf q}_\perp$ and ${\bf H}_\perp$  can either be parallel or perpendicular to each other. 
The preferred mutual orientation can  be changed  at some pressure if the orientation-dependent part of the susceptibility changes sign.
 
\section{Conclusion}

Some time ago \cite{Michal}, we  have demonstrated the   softening of spin resonance mode  in the $d$-wave superconducting CeCoIn$_5$ under  a  basal plane magnetic field at a wave vector ${\bf q}_{IC}$ that  connects  the points of  the Fermi surface with a finite gap  $\Delta({\bf k})=-\Delta({\bf k}+{\bf q}_{IC})$. 
In the strong enough field this leads to the formation of static incommensurate AF state with two possible  types of antiferromagnetic domains.
Here, we have shown that the spin-orbit interaction in a tetragonal metal
under the basal plane magnetic field acts to favor an inhomogeneous spin density  modulation 
directed  either perpendicular or parallel  to the field direction. Hence, in general, only one type of antiferromagnetic domain is energetically favorable.
This allows us to explain the puzzle of antiferromagnetic domain switching initiated by  the basal plane magnetic field rotation observed \cite{Gerber} in superconducting CeCoIn$_5$.

In CeCoIn$_5$ at ambient pressure, the incommensurate AF modulation prefers to be directed  along the direction that is '"he most perpendicular" to the field direction.  At some pressure, the preferred orientation can  change to "the most parrallel" one.

The mechanism described for the orienting influence of the magnetic field on the direction of antiferromagnetic modulation has a general character and should reveal itself in an  itinerant antiferromagnet.  The phenomenon of antiferromagnetic domain switching  is suppressed by domain pinning
and can be observable only in clean enough metals,  but not in doped antiferromagnets such as  CeRh$_{1-x}$Co$_x$In$_5$.

The susceptibility $\chi_{zz}({\bf q}_\perp, {\bf H}_\perp)$ dependence from the mutual orientation ${\bf q}_\perp$ and 
$ {\bf H}_\perp$ characterizing the intensity of spin-orbit coupling in a tetragonal material can be measured by neutron scattering.

I hope that this paper will stimulate quantitative numerical calculations of $\chi_{zz}({\bf q}_\perp, {\bf H}_\perp)$  both in the normal 
and in the superconducting states based on the real band structure of CeCoIn$_5$.

\section*{Acknowledgements}

I am grateful to S. Raymond for the interest to this study and to S. Blundell  for the help in the manuscript preparation.

\begin{figure}[p]
\includegraphics
[height=.8\textheight]
{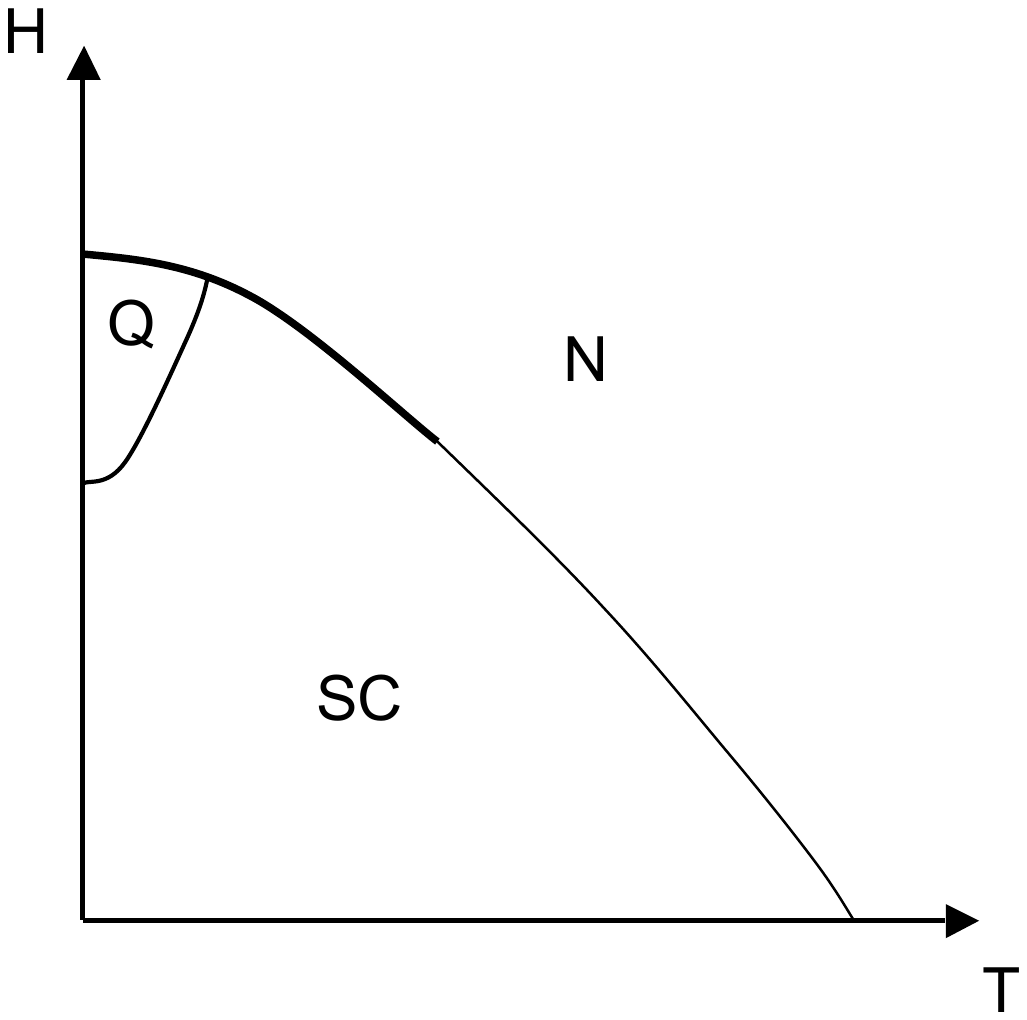}
 \caption{ Schemetical $(H,T)$ phase diagram of CeCoIn$_5$. $N$ and $SC$ are the normal and the superconducting states correspondingly. 
 The upper critical field in CeCoIn$_5$ is mostly determined by paramagnetic limiting ($H_{c2}(T=0)\simeq11.7\,\text{T}$) and due to this the phase transition to the superconducting state below $T=0.4T_c$ ($T_c=2.3\,\text{K}$)
is of the first order (thick line on the figure) \cite{FirstOrder}. The $Q$-phase is the incommensurate antiferromagnetic state coexisting with the superconducting mixed state. 
 }
\end{figure}

\begin{figure}[p]
\includegraphics
[height=.8\textheight]
{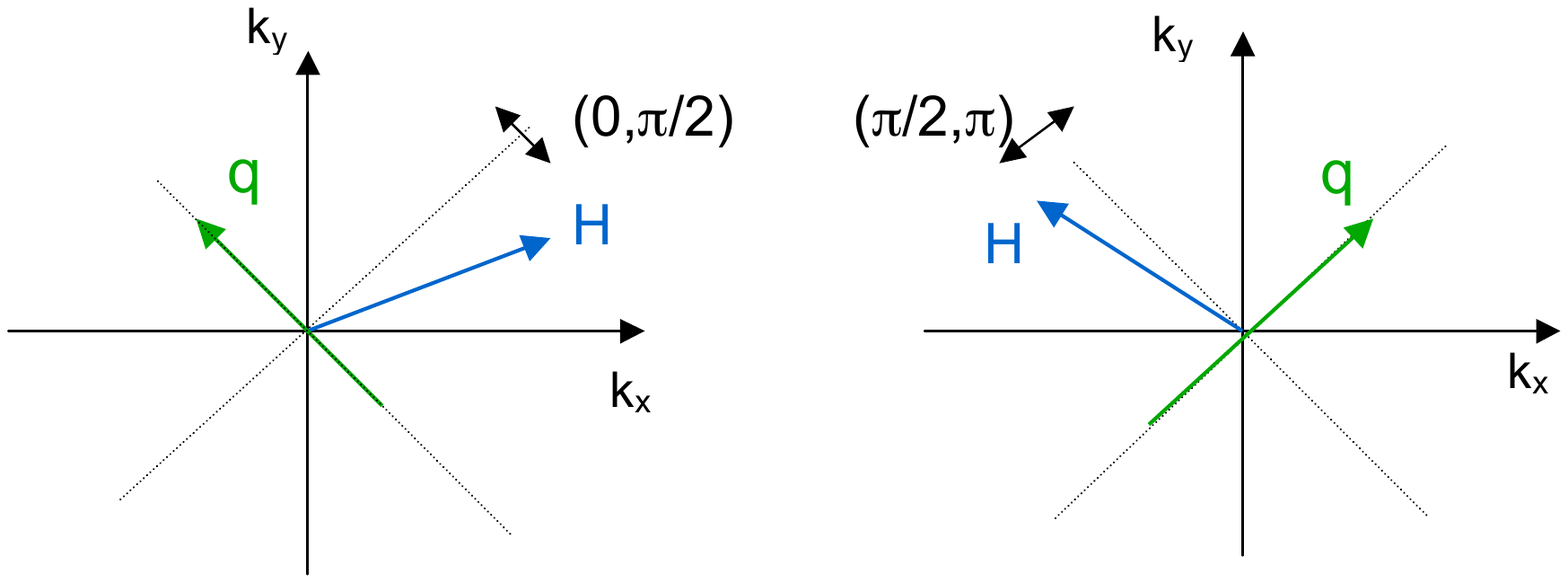}
 \caption{ (Color online).
 (Left) Basal plane magnetic field lies in the sector $(0,\frac{\pi}{2})$ or in the sector $(\pi,\frac{3\pi}{2})$, the AF modulation is directed along
$(1,\bar 1,0)$ direction. (Right)  Basal plane magnetic field lies in the sector $(\frac{\pi}{2},\pi)$ or in the sector $(\frac{3\pi}{2},2\pi)$, the AF modulation is directed along
$(1,1,0)$ direction.}
\end{figure}

\end{document}